\documentclass[12pt,noshowpacs,nofootinbib,notitlepage,amsmath]{revtex4-1}
\allowdisplaybreaks
\usepackage{graphicx,color}
\usepackage[colorlinks=true,citecolor=blue,linkcolor=blue,urlcolor=blue]{hyperref}
\usepackage[charter]{mathdesign}
\DeclareSymbolFontAlphabet{\mathcal}{symbols}
\DeclareSymbolFont{symbols}{OMS}{xmdcmsy}{m}{n}
\DeclareSymbolFont{largesymbols}{OMX}{xmdcmex}{m}{n}
\SetSymbolFont{symbols}{bold}{OMS}{xmdcmsy}{b}{n}

\begin{document}  
\title{\color{blue}\Large Dark Matter Theories in the Light of Diphoton Excess}
\author{Huayong Han}
\email{hhycqu@cqu.edu.cn}
\author{Shaoming Wang}
\email{smwang@cqu.edu.cn}
\author{Sibo Zheng}
\email{sibozheng.zju@gmail.com}
\affiliation{Department of Physics, Chongqing University, Chongqing, 401331, China}
\begin{abstract}
A new type of dark matter (DM) theories are proposed in the light of the standard model (SM) singlet scalar $\phi$ which is responsible for the diphoton excess at the LHC Run 2.
In the so-called $\phi$-portal DM models, 
after taking into account the LHC constraints and DM direct detection limits,
we show that in the perturbative framework DM as either a SM singlet scalar 
or Dirac fermion can be allowed in a wide mass range between 400 GeV and 3 TeV. 
The DM can be directly detected in SM multi-jets and missing energy.
\end{abstract}
\maketitle

\section{Introduction}
\label{intro}
Recently, a diphoton signal excess at 750 GeV was reported in the data of Large Hadron Collider (LHC) Run 2 with $pp$ collisions at energy $\sqrt{s}=13$ TeV \cite{13tevatlas,13tevcms}.
It can be explained by a SM singlet $\phi$, 
with production cross section \cite{1512.04933},
\begin{eqnarray}\label{excess}
\sigma(pp\rightarrow \phi\rightarrow \gamma\gamma)= (8\pm 3)~\text{fb}.
\end{eqnarray}
The local significance of this excess is about 3.9 $\sigma$ and 2.6 $\sigma$ for ATLAS and CMS, respectively,
and it was not seen in the data of LHC Run 1 with $\sqrt{s}=8$ TeV \cite{1506.02301, 8tev-diphoton}.
Although no excesses in $ZZ$, $WW$, $ZW$, dilepton and dijet channels \cite{8tevcms, 8tevatlas, 8tev-zz, 8tev-dijet, 8tev-bb, 8tev-tt} were observed yet 
both in the old data of Run 1 and the first data of LHC Run 2, 
this bump has stimulated great interests. 

In this paper, we consider the interest of building connections between the SM singlet scalar $\phi$
responsible for the diphoton excess and weakly-interacting massive DM models.
Inspired by the construction of SM Higgs-portal DM models \cite{Zee, 0702143, 0106249,  0003350, 0011335,1509.01765, 1510.06165},
in which the DM communicates with the SM particles via the Higgs mediator,
we  propose $\phi$-portal DM model by replacing the Higgs scalar with $\phi$.
For discussions about DM roles in diphoton excess in Eq.(\ref{excess}), 
see Refs. \cite{1512.04913,1512.04917,1512.06376, 1512.06787, 1512.06828, 1512.06828}.

Our construction of $\phi$-portal DM obviously differs from the Higgs-portal DM models
due to obviously different mediator scalar mass and Yukawa coupling.
But they indeed share a common feature, 
i.e., they are both effective theories at the electroweak (EW) scale.

This paper is organized as follows. 
In Sec.~\ref{sec2}, we define our model introduce model parameters.
In Sec.~\ref{sec3}, we consider the constraints on $\psi$-sector from diphoton excess in Eq.(\ref{excess}) and 8-TeV limits at LHC.
In particular, parameter space should be consistent with the 8-TeV $\gamma\gamma$ limit.
DM of type either a SM singlet scalar or Dirac fermion  are both addressed. 
In Sec.~\ref{sec4}, we consider the constraints on DM-sector from DM relic abundance \cite{1303.5076} and direct detection limits \cite{Xenon100, Xenon1T, LUX}.
In Sec.~\ref{sec4} we add a few comments on DM direct detection at LHC Run 2.
Finally,  we conclude in Sec.~\ref{sec5} .

\section{The Model}\label{sec2}
We assume that the effective theory of new physics model includes a scalar $\phi$ responsible for the diphoton excess, a fermion $\psi$ charged under SM gauge group $SU(3)_{c}\times U(1)_{Y}$ with electric charge $Q_{\psi}$ in unit of $e$, and a SM singlet scalar DM ($\phi_{\text{DM}}$) or a SM singlet fermion DM ($\psi_{\text{DM}}$).
The TeV-scale effective Lagrangian $\mathcal{L}_{\text{eff}}$ for this model  is given by,
\begin{eqnarray}\label{Lag}
\mathcal{L}_{\text{eff}}=\mathcal{L}_{\phi}+\mathcal{L}_{\psi}+\mathcal{L}_{\text{DM}}
+\mathcal{L}_{\text{Yukawa}}
\end{eqnarray}
where
\begin{eqnarray}\label{interaction}
\mathcal{L}_{\phi}&=& \frac{1}{2} \partial_{\mu}\phi\partial^{\mu}\phi -\frac{1}{2}m_{\phi}^{2}\phi^{2},\nonumber\\
\mathcal{L}_{\psi}&=&i\bar{\psi}\gamma^{\mu}D_{\mu}\psi -m_{\psi}\bar{\psi}\psi, \nonumber\\
\mathcal{L}_{\text{DM}}&=& \begin{cases}
\frac{1}{2} \partial_{\mu}\phi_{\text{DM}}\partial^{\mu}\phi_{\text{DM}} -\frac{1}{2}m^{2}_{\text{DM}}\phi^{2}_{\text{DM}}, &  (\text{scalar}) \\
i\bar{\psi}_{\text{DM}}\gamma^{\mu}D_{\mu}\psi_{\text{DM}}-m_{\text{DM}}\bar{\psi}_{\text{DM}}\psi_{\text{DM}} , &  (\text{fermion})
\end{cases} \nonumber\\
\mathcal{L}_{\text{Yukawa}}&=&y\phi \bar{\psi} \psi+
\begin{cases}
\kappa \upsilon_{EW} \phi \phi_{\text{DM}}\phi_{\text{DM}}, &  (\text{scalar}) \\
\kappa \phi \bar{\psi}_{\text{DM}} \psi_{\text{DM}}, &  (\text{fermion}).
\end{cases}
\end{eqnarray} 
We identify $m_{\phi}=750$ GeV and EW scale $\upsilon_{\text{EW}}=246$ GeV.
The Yukawa coupling $\kappa$ in the case for scalar DM has been normalized to a dimensionless parameter.
For either a scalar or fermion DM there are four following model parameters in Eq.(\ref{Lag}),
\begin{eqnarray}\label{ps}
\{y,~m_{\psi}; \kappa,~m_{\text{DM}}\}
\end{eqnarray}
Similar to Higgs-portal singlet scalar DM \cite{Zee, 0702143, 0106249,  0003350, 0011335}, a $Z_{2}$ parity,  
under which $\phi_{\text{DM}}$ is odd and the others are even,
is employed to keep DM stable.

Instead of writing the interactions between $\phi$ and gluons and photons via operators with mass dimension five, 
in (\ref{Lag}) we consider an explicit realization via fermion $\psi$ 
\footnote{ One may care about the problem of gauge anomaly,
which can be avoided by embedding $\psi$ into vector-like quark models.
We temporarily assume that those fermions ignored here are not relevant for our study.}.
The advantage is obvious, as the number of model parameters are reduced. See, e.g., \cite{1512.06562}.

\begin{figure}[!h]
\includegraphics[width=4in]{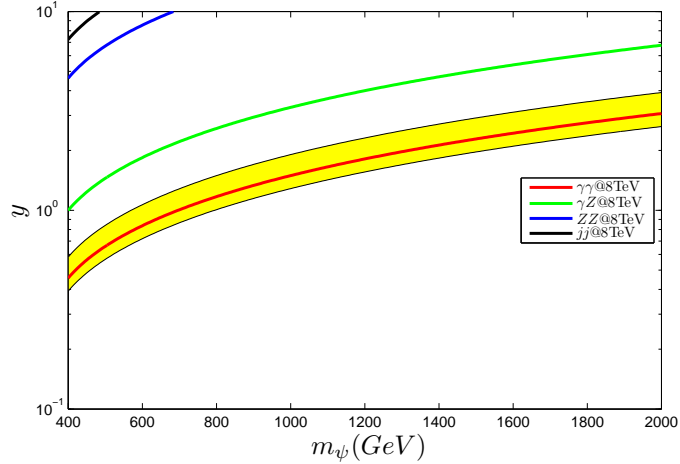}
\caption{Yellow band corresponds to contours of observed diphoton signal strength in the parameter space for $Q_{\psi}=5/3$. 
Limits in the data of LHC Run 1 are shown in curves,
above which regions are excluded.}
\label{signal}
\end{figure}

\section{Constraints on $\psi$-Sector in Diphoton Excess}\label{sec3}
The decay channels for $\phi$ include $\phi\rightarrow \{gg,\gamma\gamma,ZZ, Z\gamma, \}$ and an extra one $\phi\rightarrow \bar{\psi}\psi$ in the low mass region $m_{\psi}< m_{\phi}/2 $.
Some of these magnitudes satisfy
\begin{eqnarray}\label{width}
\frac{\Gamma (\phi\rightarrow \gamma\gamma)}{\Gamma (\phi\rightarrow gg)} \propto Q^{4}_{\psi} ,
\end{eqnarray}
while the others such as $\Gamma (\phi\rightarrow \{\gamma\gamma,Z\gamma,ZZ\})$
are all of the same order.
Obviously, larger branching ratio $\text{Br}(\phi\rightarrow \gamma\gamma)$ 
in Eq.(\ref{width}) can be obtained by choosing larger $Q_{\psi}$ .
In this paper we simply choose the electric charge 5/3 for $\psi$ for the following facts.  
At first, the dark matter phenomenology discussed in Sec. IV does not significantly affected by the choice on electric charge slightly bigger  than $5/3$ for $\psi$ as favored by the diphoton excess at 750 GeV. 
Secondly,  those results arising from electric charge $5/3$ in Sec.IV are useful reference for the case where $\psi$ is also charged under $SU(2) _L$\footnote{Being $SU(2)_L$ singlet, the electric charge for $\psi$ can be alternatively chosen bigger than $5/3$. See our previous work \cite{1512.06562} for relevant discussion.
In contrast, if  $\psi$ is further charged under $SU(2)_L$,  
the mass region $m_{\psi}<800$ GeV has been excluded by the 8-TeV LHC data \cite{1312.2391} for the assumption $\text{Br}(\psi\rightarrow tW)=100\%$. 
Similarly, see Ref. \cite{1311.7667} for LHC lower mass bounds for $Q_{\psi}=2/3$.}. 
In the setting on small other than large decay width for $\phi$, 
$m_{\psi}$ is restricted to be above $m_{\phi}/2$.

The production cross section $\sigma(pp\rightarrow \phi)$ is mainly through gluon fusion, the magnitude of which depends on Yukawa coupling $y$ and fermion mass $m_{\psi}$.
Given the fact that the decay width ratio $\Gamma(\phi\rightarrow \gamma\gamma)/\Gamma(\phi\rightarrow gg)$ is nearly fixed for the explicit choice on electric charge $Q_{\psi}$,
the diphoton signal strength $\sigma(pp\rightarrow\phi\rightarrow\gamma\gamma)$ is sensitive to $y$ and fermion mass $m_{\psi}$.
In Fig.\ref{signal} the yellow band corresponds to the contours of observed diphoton signal strength, with limits in the data of LHC Run 1 are shown in curves. 
It indicates that the narrow region in the yellow band below the limit given by $\gamma\gamma@8$TeV can explain the diphoton excess 
and is consistent with the following limits \cite{1506.02301,8tevcms, 8tevatlas,
8tev-diphoton, 8tev-zz, 8tev-dijet, 8tev-bb, 8tev-tt} at $8$-TeV LHC simultaneously, 
\begin{eqnarray}\label{limits}
\sigma(pp\rightarrow \gamma\gamma) &<& 1.5~\text{fb},~~(\text{red}), \nonumber\\
\sigma(pp\rightarrow Z\gamma) &<& 4~\text{fb}, ~~(\text{green}),\nonumber\\
\sigma(pp\rightarrow ZZ) &<& 12~\text{fb}, ~~(\text{blue}),\nonumber\\
\sigma(pp\rightarrow jj) &<& 2.5~\text{pb}, ~~(\text{black}),
\end{eqnarray}
It also implies that the value of $y$ as required is in the perturbative region for $m_{\psi}$ below 2 TeV.

\begin{figure}
\centering
\begin{minipage}[b]{0.5\textwidth}
\centering
\includegraphics[width=2.8in]{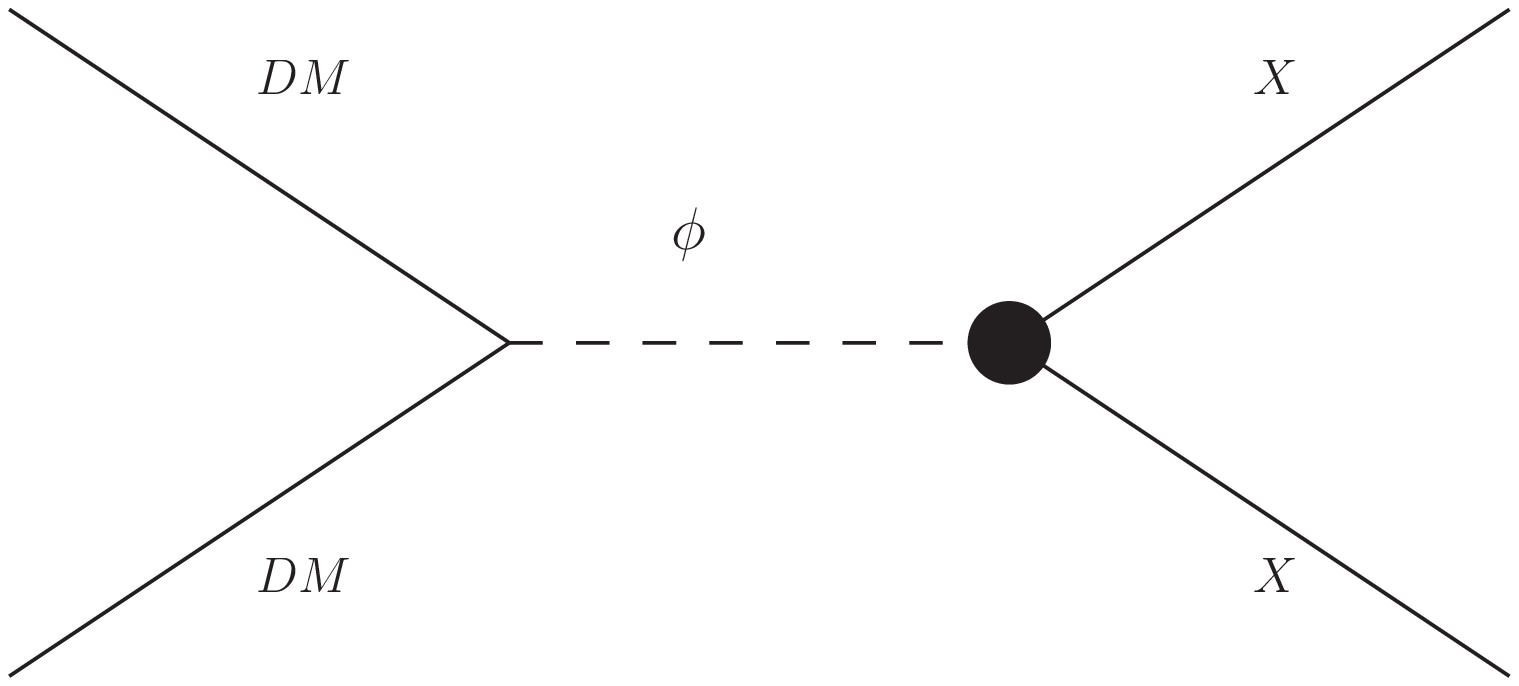}
\end{minipage}%
\centering
\begin{minipage}[b]{0.5\textwidth}
\centering
\includegraphics[width=2.8in]{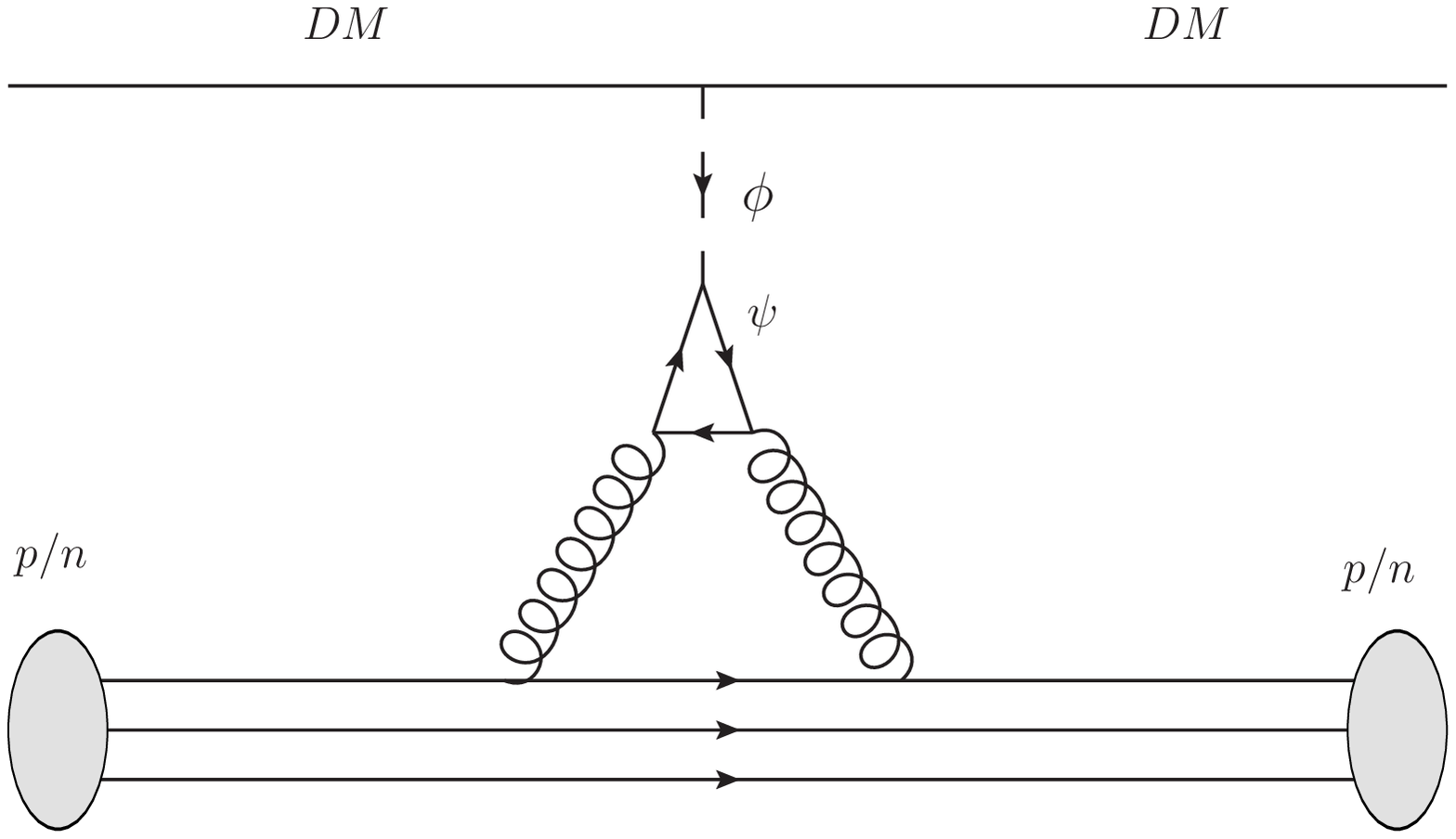}
\end{minipage}%
\caption{$\mathbf{Left~panel}$: Feynman diagrams for DM annihilation. 
$\mathbf{Right~panel}$: Feynman diagram for DM-nucleon scattering cross section.}
\label{feynman}
\end{figure}

\section{Constraints on DM-Sector}\label{sec4}
In this section we take the DM view on our model parameters.
The Yukawa interaction between DM and $\phi$ responsible for diphoton excess provides  the $s$-channel annihilation 
\begin{eqnarray}\label{relic}
\phi_{\text{DM}}\phi_{\text{DM}}\rightarrow \{XX\},  ~~~~X=\{\psi,\phi, \text{SM}\},
\end{eqnarray} 
as shown in the left panel of Fig.\ref{feynman}.
This annihilation should account for the DM relic abundance as measured 
by the Plank and WMAP 9-year data \cite{1303.5076},
\begin{eqnarray}\label{relic}
\Omega_{DM} h^{2}=0.1199\pm 0.0027.
\end{eqnarray} 

By employing micrOMEGAs \cite{1407.6129} we show in Fig.\ref{relic} the contours of $\Omega_{DM}h^{2}$ in the parameter space of $\kappa$ and $m_{DM}$ with DM either a scalar or Dirac fermion for three representative choices on $m_{\psi}=\{400, 800, 1000, 2000\}$ GeV.
$\mathbf{Left~panel}$ in Fig.\ref{relic} shows that once $\phi_{\text{DM}}\phi_{\text{DM}}\rightarrow \psi\bar{\psi}$ is opened in the mass region  $m_{\phi_{\text{DM}}}>m_{\psi}$, 
annihilation cross section significantly increases,
which leads to smaller $\kappa$ as required by the relic abundance in Eq.(\ref{relic}). 
Similar phenomenon occurs in the $\mathbf{Right~panel}$ for Dirac fermion DM,
although it changes more mildly.
This difference may be due to the particular normalization we have chosen for scalar DM in Eq.(\ref{Lag}).
The requirement of dark matter relic density as indicated by the right panel of Fig.3 implies that perturbative analysis is not valid for fermion DM with mass roughly below 400 GeV. Conversely, this requirement as indicated by the left panel of Fig.3  implies that perturbative analysis is not valid for scalar DM with mass roughly above 2.5 TeV instead.

Unlike the Higgs-portal singlet DM model where DM-nucleon scattering proceeds via tree-level process, ,
in our model DM-nucleon scattering proceeds via a loop process instead,
as shown in the $\mathbf{right~panel}$ of Fig.\ref{feynman},
where the intermediate vector boson in the $\mathbf{right~panel}$ is either SM gluon or photon.
As a result, the $\phi$-nucleon scattering cross section $\sigma_{\text{SI}}$ 
is relatively suppressed in compared with a tree-level process.
Fig.\ref{si} shows the contours of $\sigma_{\text{SI}}$ as function of DM mass
for either scalar ($\mathbf{left~panel}$)or Dirac fermion ($\mathbf{right~panel}$) DM, which clearly shows that it is consistent with all present direct detection limits.
It also implies that there is no prospect for discovery even in the further experiment at Xenon1T.

\begin{figure}
\centering
\begin{minipage}[b]{0.5\textwidth}
\centering
\includegraphics[width=3.2in]{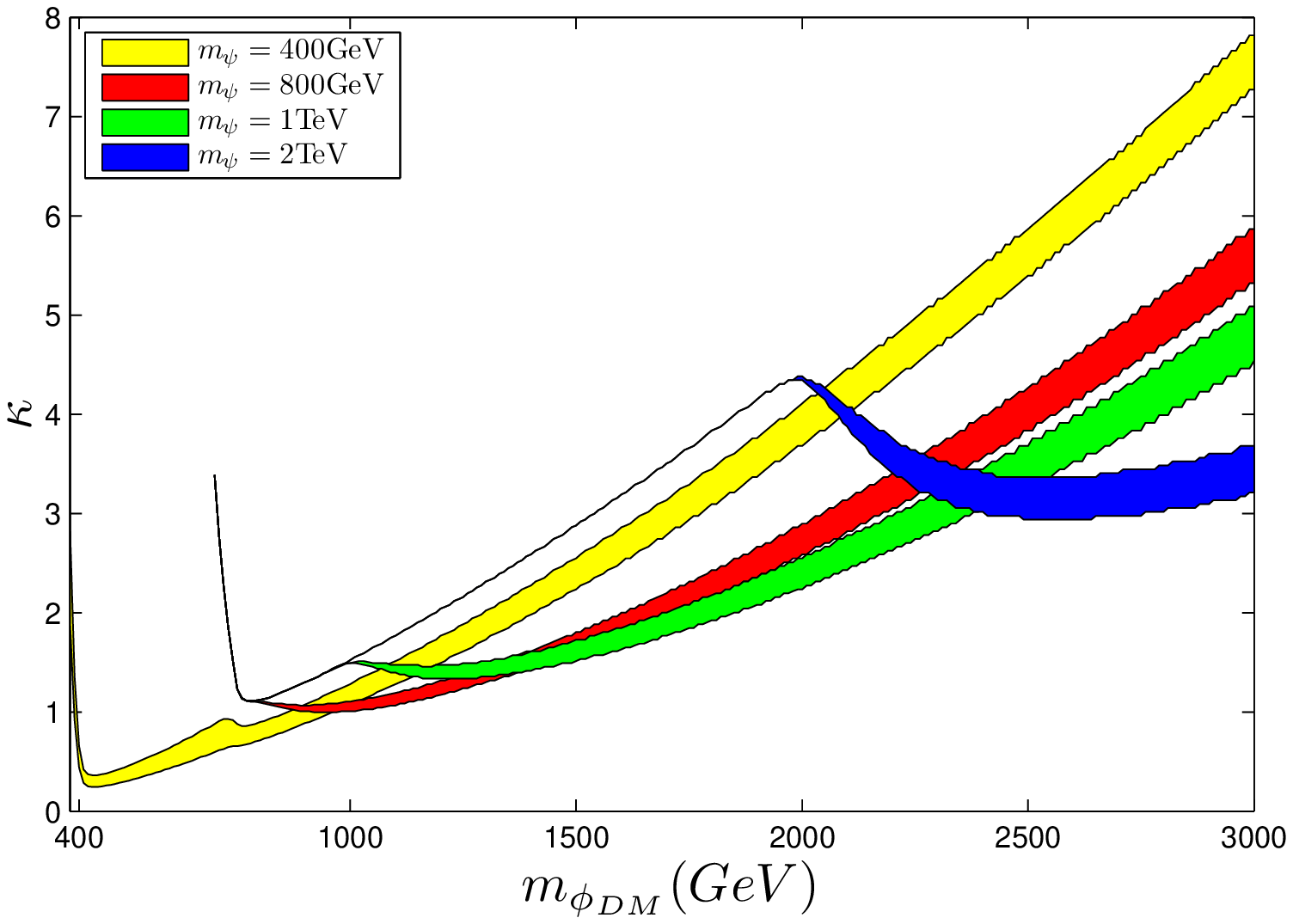}
\end{minipage}%
\centering
\begin{minipage}[b]{0.5\textwidth}
\centering
\includegraphics[width=3.2in]{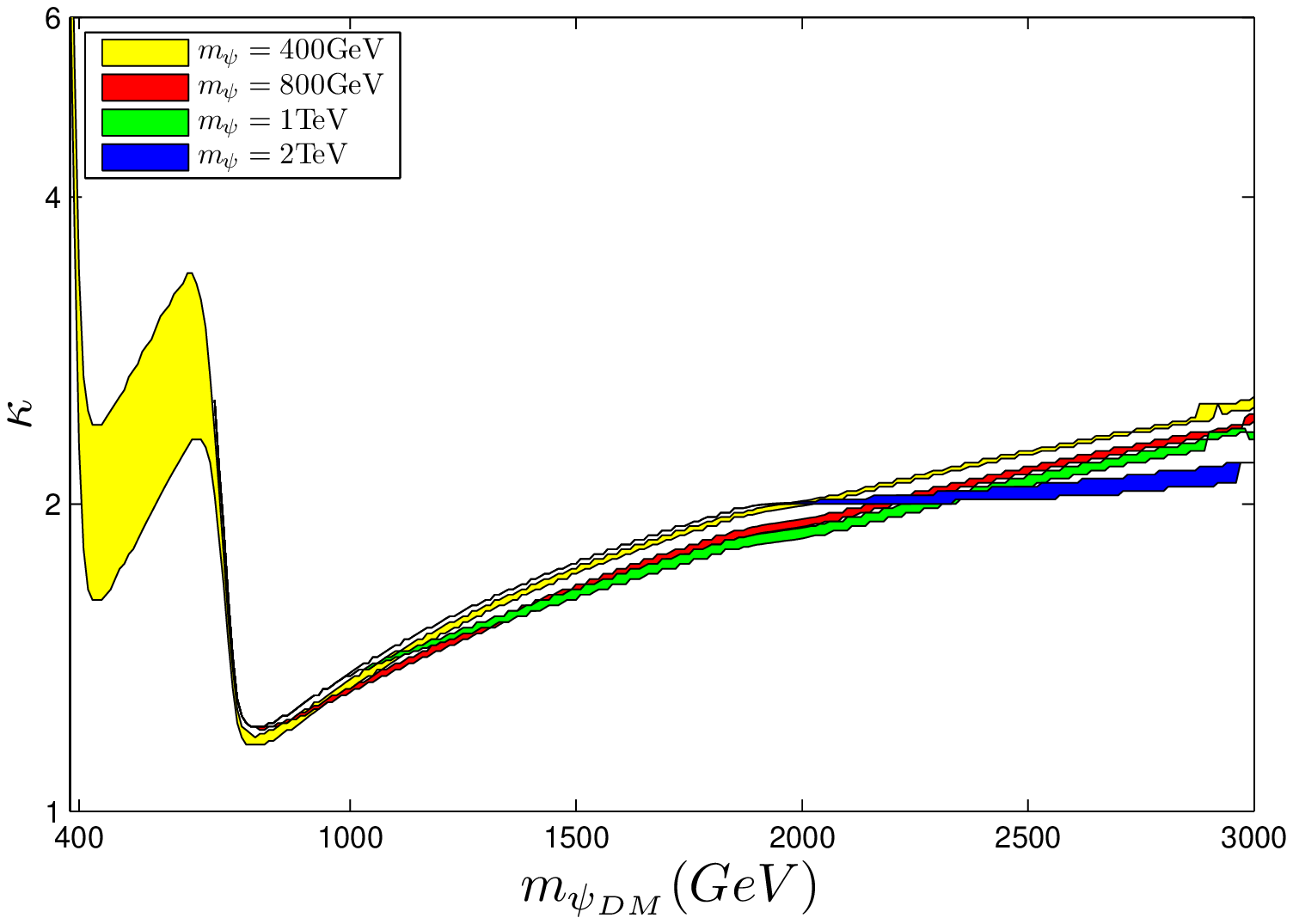}
\end{minipage}%
\caption{Contours of $\Omega_{DM}h^{2}$ projected to the parameter space of $\kappa$ and $m_{DM}$.
$\mathbf{Left~panel}$ and $\mathbf{Right~panel}$ corresponds to a scalar and Dirac fermion DM, respectively.
In each case, we take three representative choices on $m_{\psi}=\{400, 800, 1000, 2000\}$ GeV. The uncertainty for $\kappa$ is due to the uncertainty of $y$ as shown in Fig.\ref{signal}.}
\label{relic}
\end{figure}

\begin{figure}
\centering
\begin{minipage}[b]{0.5\textwidth}
\centering
\includegraphics[width=3.2in]{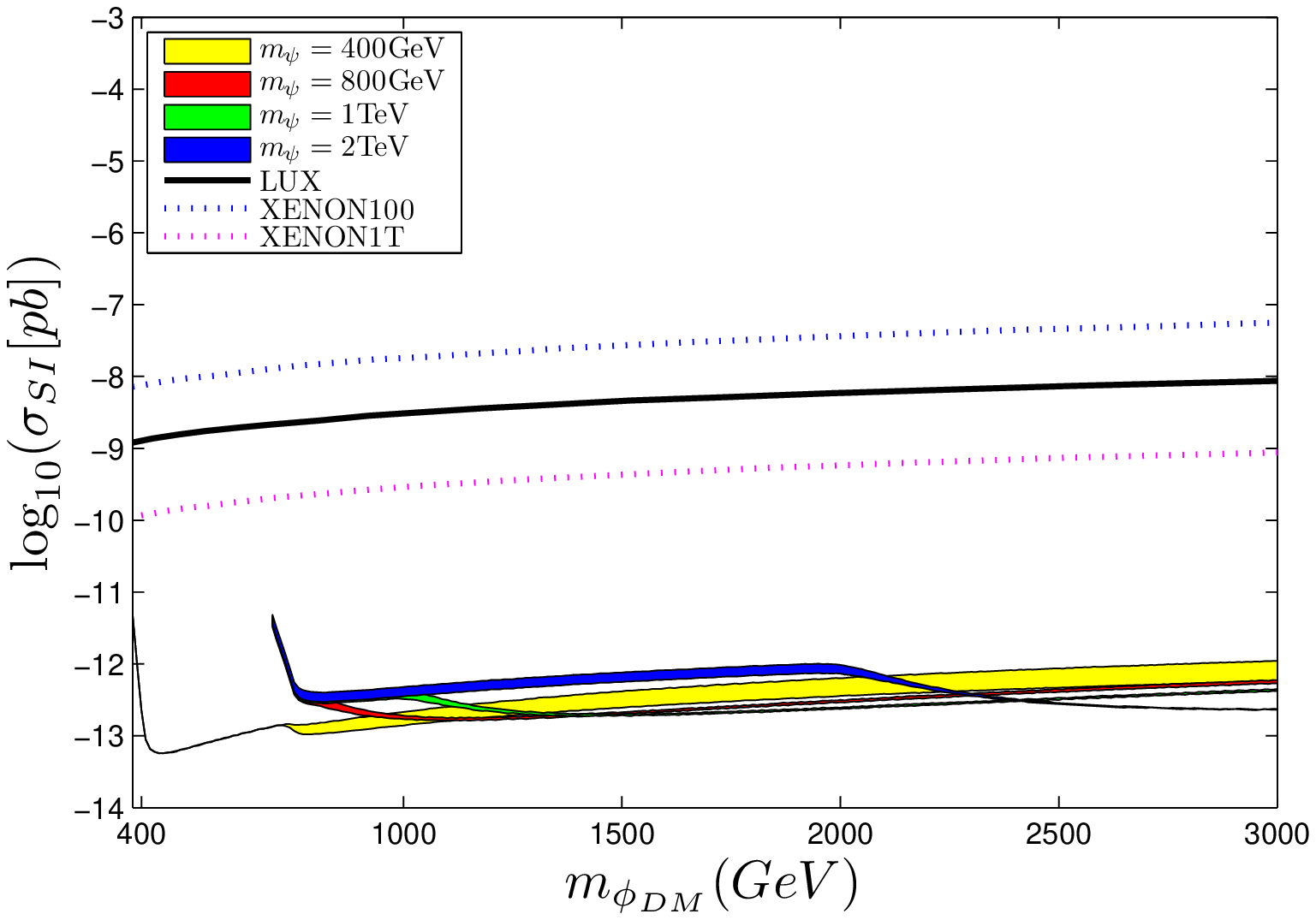}
\end{minipage}%
\centering
\begin{minipage}[b]{0.5\textwidth}
\centering
\includegraphics[width=3.2in]{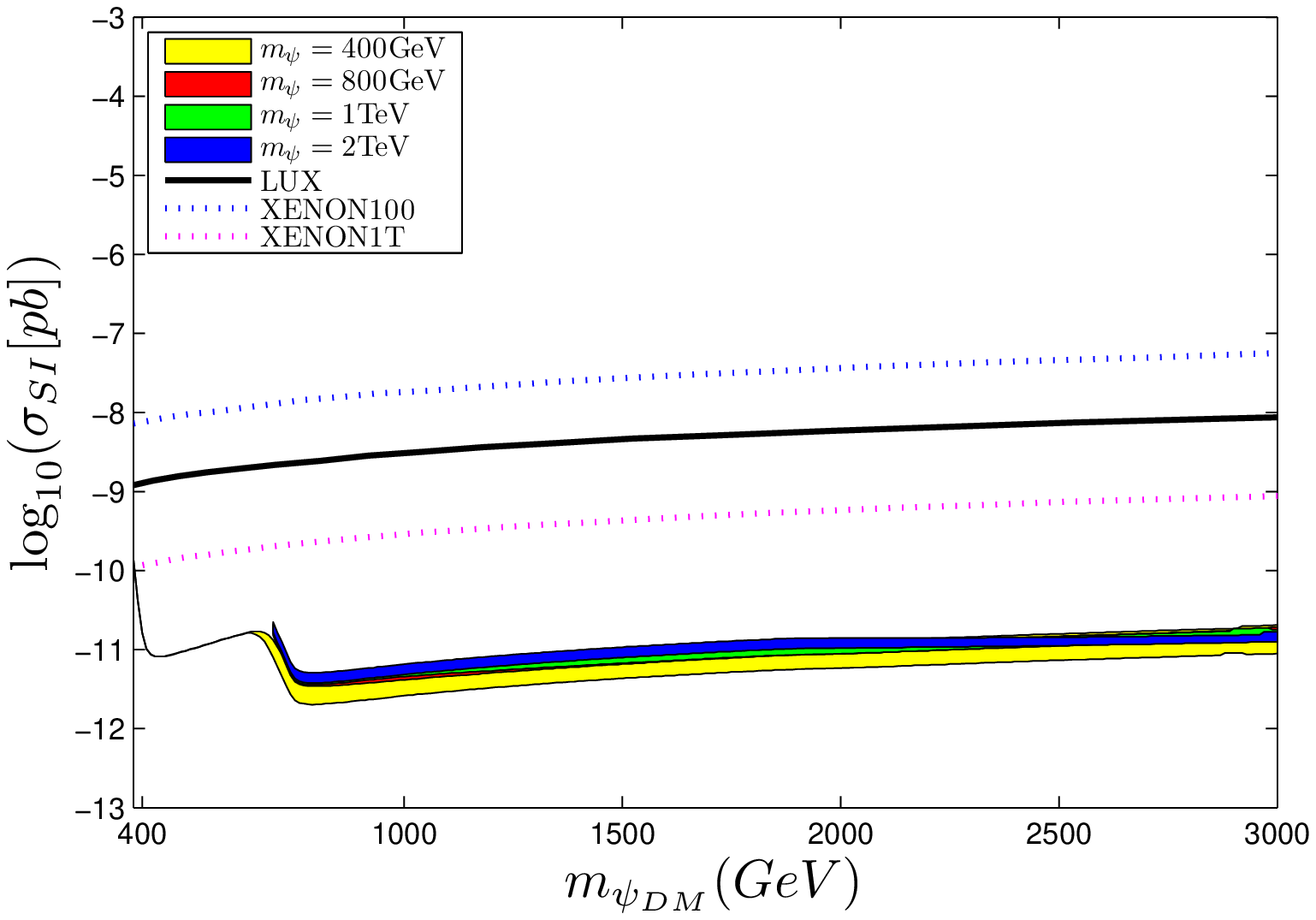}
\end{minipage}%
\caption{Contours of $\sigma_{\text{SI}}$ as function of DM mass
for either scalar ($\mathbf{left~panel}$)or Dirac fermion ($\mathbf{right~panel}$) DM
for three representative choices on $m_{\psi}=\{400, 800, 1000, 2000\}$ GeV, 
which clearly show that our DM model is consistent with all present direct detection limits.
They also indicate that there is no prospect for discovery at further Xenon1T experiment.
Similar to Fig.\ref{relic},
the uncertainty for $\kappa$ is due to the uncertainty of $y$ in Fig.\ref{signal}.}
\label{si}
\end{figure}

\section{DM $@$ LHC Run 2}\label{sec4}
The DM can be directly detected at LHC via DM pair production.
It may contribute to excess in SM multi-jets plus missing energy.
The signal strength of multi-jet process $\sigma(pp\rightarrow\phi+X\rightarrow\text{X+DM~DM})$ through gluon fusion
 is suppressed by many-body final states.
Consider that the DM is typically in the mass range between 400 GeV and 3 TeV
and the SM background is very large, 
a large luminosity is required for discovery of DM at the LHC Run 2.
This issue will be addressed elsewhere in detail \cite{Han:2016}.

\section{Conclusions}\label{sec5}
In this paper we propose a type of new DM models, 
in which the scalar responsible for the diphoton excess at 13-TeV LHC mediates the 
interactions between DM and SM.
In the so-called $\phi$-portal DM models, 
after taking into account the LHC constraints and DM direct detection limits,
we show that in the perturbative framework DM either a SM singlet scalar or Dirac fermion can be allowed in a wide mass range between 400 GeV and 3 TeV. 
With high integrated luminosity, the DM can be directly detected in SM multi-jets and missing energy.

\begin{acknowledgments}
This work is supported in part by the Natural Science Foundation of China under grant No.11205255 (S. W) and 11405015 (S. Z).
\end{acknowledgments}

\linespread{1}

\end{document}